\documentclass[x11names]{llncs}
\usepackage[utf8]{inputenc} %
\usepackage[OT1]{fontenc}
\usepackage{amsfonts,listings,courier,pgfplotstable,times,xspace,tikz,amssymb,
  amsmath,comment,stmaryrd,subcaption,multirow,wrapfig,pifont,bigdelim}
\usetikzlibrary{shapes,arrows,chains,fit,backgrounds}

\tikzstyle{bef}=[align=left,font=\ttfamily,node distance=3mm,text width=\textwidth*.4]
\tikzstyle{aft}=[align=left,font=\ttfamily,node distance=3mm,text width=\textwidth*.5]
\tikzstyle{acsl}=[inner sep=0mm,fill=blue!15]
\tikzstyle{genc}=[inner sep=0mm,fill=blue!15]

\tikzstyle{common}=[node distance=15mm,align=center]
\tikzstyle{test}=[draw,trapezium,trapezium left angle=70,
  trapezium right angle=-70,common]
\tikzstyle{op}=[draw,common]
\tikzstyle{data}=[draw,common,ellipse,node distance=35mm]
\tikzstyle{ce}=[data]
\tikzstyle{arrow}=[draw,->]
\tikzstyle{darrow}=[draw,o->]

\newcommand{\toolname}{\textsc{RPP}\xspace}
\renewcommand{\phi}{\varphi}

\newcommand{\framac}{\textsc{Frama-C}\xspace}

\newcommand{\acsl}{\textsc{acsl}\xspace}

\newcommand{\Wp}{\textsc{Wp}\xspace}

\lstdefinelanguage{pretty-ACSL}{%
  escapechar={},
  literate=
   {==}{{$==$}}2
   {==>}{{$\Rightarrow$}}1
   {integer\ i}{{i$\,\in \mathbb{Z}\,$}}4
   {integer\ j}{{j$\,\in \mathbb{Z}\,$}}4
   {integer\ k}{{k$\,\in \mathbb{Z}\,$}}4
   {integer\ m}{{m$\,\in \mathbb{Z}\,$}}4
   {integer\ l}{{l$\,\in \mathbb{Z}\,$}}4
   {\\forall}{{$\forall$}}1
   {\\exists}{{$\exists$}}1
   {integer}{{$\mathbb{Z}$}}1
   {real}{{$\mathbb{R}$}}1
   {&&}{{$\wedge$}}1
   {||}{{$\vee$}}1
   {!=}{{$\neq$}}1
   {<}{{$<$}}1
   {<=}{{$\le~$}}1
   {>}{{$>$}}1
   {>=}{{$\ge~$}}1
   {<==>}{{$\Leftrightarrow$}}1,
  morekeywords={assert,assigns,assumes,axiom,axiomatic,behavior,behaviors,
    boolean,breaks,complete,continues,data,decreases,disjoint,ensures,
    exit_behavior,ghost,global,inductive,invariant,lemma,logic,loop,
    model,predicate,relational,reads,requires,sizeof,strong,struct,terminates,
    type,union,variant,uchar,byte,typically,\\result,\\old,\\at,\\valid,
    \\separated,\\nothing,Pre,\\sum,\\numof,\\call,\\from},
  alsoletter={\\},
  morecomment=[l]{//}
}
\lstnewenvironment{pretty-codeACSL}{\lstset{language=pretty-ACSL,stepnumber=0}}{\smallskip}

\lstdefinelanguage{ACSL}{%
  escapechar={},
  literate=,
  morekeywords={assert,assigns,assumes,axiom,axiomatic,behavior,behaviors,
    boolean,breaks,complete,continues,data,decreases,disjoint,ensures,
    exit_behavior,ghost,global,inductive,invariant,lemma,logic,loop,
    model,relational,predicate,reads,requires,sizeof,strong,struct,terminates,
    type,union,variant,uchar,byte,typically,\\result,\\old,\\at,\\valid,
    \\separated,\\nothing,Pre,\\exists,\\forall,\\sum,\\numof},
  alsoletter={\\},
  morecomment=[l]{//}
}
\lstnewenvironment{codeACSL}{\lstset{language=ACSL,stepnumber=0}}{\smallskip}

\lstdefinestyle{pretty-c}{language={[ANSI]C},%
  alsolanguage=pretty-ACSL,%
  moredelim={*[l]{//}},%
  deletecomment={[s]{/*}{*/}},
  moredelim={*[l]{//@}},%
}

\lstdefinestyle{c}{language={[ANSI]C},%
  alsolanguage=ACSL,%
  moredelim={*[l]{//}},%
  deletecomment={[s]{/*}{*/}},
  moredelim={*[l]{//@}},%
}

\newcolumntype{C}[1]{>{\centering\let\newline\\\arraybackslash\hspace{0pt}}m{#1}}

\begin{document}

\pagestyle{headings}

\title{\toolname: Automatic Proof of Relational Properties by Self-Composition}
\author{
  \mbox{}
  \hspace{-4mm}
  Lionel Blatter\inst{1} \and
  Nikolai Kosmatov\inst{1} \and
  Pascale Le Gall\inst{2}  \and
  Virgile Prevosto\inst{1}
}
\institute{
  CEA, LIST, Software Reliability and Security Laboratory,
  91191 Gif-sur-Yvette France\\
  \email{firstname.lastname@cea.fr}
  \and
  Laboratoire de Mathématiques et Informatique pour la Complexité et les Systèmes\\
  CentraleSupélec, Université Paris-Saclay, 92295 Châtenay-Malabry France\\
  \email{firstname.lastname@centralesupelec.fr}
}

\maketitle
\vspace{-3mm}
\begin{abstract}

  Self-composition provides a powerful theoretical approach to prove relational properties,
  i.e. properties relating several program executions,
  that has been applied to compare two runs of one or similar programs
  (in secure dataflow properties, code transformations, etc.).
  This tool demo paper presents \toolname, an original implementation of self-composition for specification 
  and verification of relational properties in C programs in the \framac platform.
  We consider a very general notion of relational properties invoking any finite number of 
  function calls of possibly dissimilar functions with possible nested calls.
  The new tool allows the user to specify a relational property,
  to prove it in a completely automatic way using classic deductive
  verification, and
  to use it as a hypothesis in the proof of other properties that may rely on it.

\smallskip

\end{abstract}
\vspace{-4mm}

\vspace{-3mm}
\section{Introduction}
\vspace{-3mm}
\label{sec:intro}

Modular deductive verification 
allows the user to prove that a function respects its formal specification.
For a given function $f$, any individual call to $f$ can be proved to respect the \emph{contract} of $f$,
that is, basically an implication: if the given \emph{precondition} is true before the call, 
the given \emph{postcondition} is true after it.
However, some kinds of properties
are not reduced to one function call.
Indeed, 
it is frequently necessary to express a property that involves several functions
or relates the results of several calls to the same function for different arguments. 
We call them \emph{relational properties}.

Different theories and techniques have been proposed to deal with relational properties in different contexts.
They include Relational Hoare Logic 
to show the equivalence of program transformations \cite{DBLP:conf/popl/Benton04} 
or Cartesian Hoare Logic for  $k$-safety properties \cite{SousaD16}.
Self-composition  \cite{DBLP:journals/mscs/BartheDR11} is a 
theoretical approach to prove relational properties relating two execution traces.
It reduces the verification of a relational property 
to a standard verification problem of a new function. 
Self-composition techniques have been applied for verification of information flow properties 
\cite{DBLP:journals/mscs/BartheDR11,prograprodu}
and properties of two equivalent-result object methods \cite{DBLP:conf/esop/LeinoM08}.
Relational properties can be expressed on 
Java pure methods \cite{JMLfunctionCall} using the JML specification language. 
OpenJML \cite{cok14} offers a partial support for deductive verification of relational properties.
The purpose of the present work is to implement and extend self-composition for specification 
and verification of relational properties in the context of
the \acsl specification language~\cite{ACSL} and the deductive 
verification plugin \Wp of  \framac \cite{Frama-C}.
We consider a large class of relational properties (universally quantified properties
invoking any finite number of 
calls of possibly dissimilar functions with possibly nested calls),
and propose an automatic solution allowing the user not 
only to prove a relational property, but also 
to use it as a hypothesis.


\smallskip\noindent
\textbf{Motivation.} 
The necessity to deal with relational properties in  \framac has been faced in various verification projects. 
Recent work \cite{BishopBC13} reports on verification of continuous monotonic functions
in an industrial case study on smart sensor software. The authors write:
``After reviewing around twenty possible code analysis tools, 
we decided to use \framac, which fulfilled all our 
requirements (apart from the specifications involving the 
comparison of function calls).'' 
The relational property in question is the monotonicity of a function
(e.g., \smash{$x \leq y \Rightarrow f(x) \leq f(y)$}).
To deal with it in \framac,
\cite{BishopBC13}
applies a variation of self-composition
consisting in a separate verification of an additional,
manually created wrapper function simulating the calls to be compared.

Relational properties can often be useful to give an expressive specification of
library functions or hardware-supported functions, when the source code is not available.
In this case, relational properties are only specified and used to verify client code, but are not verified themselves.
For instance, 
in the PISCO project\footnote{\url{http://www.systematic-paris-region.org/en/projets/pisco}}, an industrial case study on verification of
software using hardware-provided cryptographic primitives (PKCS\#11 standard)
required tying together different functions with properties such as
\smash{$\mathrm{Decrypt}(\mathrm{Encrypt}(Msg,PrivKey),PubKey)=Msg$}.
Other examples include properties of data structures,
such as matrix transformations (e.g. $(A+B)^\intercal = A^\intercal+B^\intercal$ or
$\mathrm{det}(A) = \mathrm{det}(A^\intercal)$), 
the specification of
$\mathrm{Push}$ and $\mathrm{Pop}$ over a stack~\cite{ACSLbyExample},
or parallel program specification 
(e.g., 
\smash{$\mathrm{map}(\mathrm{append}(l_1,l_2)) = \mathrm{append}(\mathrm{map}(l_1),\mathrm{map}(l_2))$}
in the MapReduce approach).
A subclass of relational properties, \emph{metamorphic properties}, relating multiple executions 
of the same function \cite{HuiHuang2013}, are also used in a different context 
in order to address the oracle problem in software testing \cite{Weyuker82}.


Manual application of self-composition or possible workarounds reduce the level of automation, 
can be error-prone and do not provide a complete automated link 
between three key components: \textsl{(i)}~the property specification, \textsl{(ii)}~its proof, 
and \textsl{(iii)}~its usage as a hypothesis in other proofs.
Thus, the lack of support for relational properties can be a major obstacle to
a wider application of deductive verification in academic and industrial projects.
\smallskip\noindent
\textbf{The contributions} 
of this tool demo paper include:
\vspace{-2mm}
\begin{itemize}
\item a new specification mechanism to formally express a relational property in \acsl;
\item a fully-automated transformation into ACSL-annotated C code based on (an extension of) self-compo\-sition,
that allows the user to prove such a property; 
\item a generation of an axiomatic definition and additional annotations that allow us
to use a relational property 
as a hypothesis for the proof of other properties in a completely automatic and transparent way;
\item an extension of self-composition to a large class of relational properties, including several calls of possibly 
dissimilar functions and possibly nested calls, and 
\item an implementation of this approach in a \framac plugin \toolname
with a sound integration of proof statuses of relational properties. 
\end{itemize}
\vspace{-2mm}


\vspace{-3mm}
\section{The Method and the Tool}
\vspace{-3mm}
\label{subsec:example}

%
%
%

\subsection{Specification and Preprocessing of a Relational Property}
\vspace{-2mm}
\label{subsec:transformation}
The proposed solution is designed and implemented
on top of 
\framac~\cite{Frama-C}, a framework for analysis of C code developed at CEA LIST. 
\framac offers a specification language, called \acsl~\cite{ACSL}, and a deductive verification plugin, \Wp~\cite{WP}, 
that allow the user to specify the desired program properties as function contracts 
and to prove them.  
A typical \acsl function contract may include a precondition (\lstinline'requires' clause stating 
a property supposed to hold before the function call) and
a postcondition  (\lstinline'ensures' clause that should hold after the call), 
as well as a frame rule (\lstinline'assigns' clause indicating which parts
 of the global program state the function
is allowed to modify). 
An assertion (\lstinline{assert} clause) can also specify a local property at any function statement.

\smallskip\noindent
\textbf{Specification.} To specify a relational property,
we propose an extension of \acsl specification language
with a new clause, \lstinline{relational}. For technical, \framac-related,
reasons, these clauses must be attached to a function contract. Thus, a property
relating calls of different functions, such as \lstinline|R3| in
Figure \ref{fig:ex1}a, must appear in the contract of the last function involved
in the property, {\it i.e.} when all relevant functions are in scope.
To refer to several function calls  in such a property, 
we introduce a new construct \lstinline{\call(f,<args>)} 
used to indicate the value returned by the call \lstinline{f(<args>)} to \lstinline{f}
with arguments \lstinline{<args>}. 
\lstinline{\call} can be used recursively, i.e. a parameter of a called function can be the result of another function call.
For example, properties \lstinline{R1,R2} at lines 2--3, 10--11 of Figure \ref{fig:ex1}a specify monotonicity of functions \lstinline{f1,f2}, while property
R3 at line 12-13 indicates that \lstinline|f1(x)| is always less than
\lstinline|f2(x)|.


\begin{figure}[tb]
  \lstset{basicstyle=\scriptsize\ttfamily,mathescape=true}
  \begin{subfigure}{0.52\textwidth}
\begin{lstlisting}[firstline=1, lastline=17]
/*@ assigns \nothing;
  relational R1: \forall int x1,x2; 
    x1 < x2 ==> \call(f1,x1) < \call(f1,x2);
*/
int f1(int x){
  return x + 1;
}

/*@ assigns \nothing;
  relational R2: \forall int x1, x2; 
    x1 < x2 ==> \call(f2,x1) < \call(f2,x2);
  relational R3: \forall int k; 
    \call(f1,k) < \call(f2,k);
*/
int f2(int y){
  return y + 2;
}
\end{lstlisting}
    \vspace{-1mm}
  \end{subfigure}
  \hspace{2mm}
  \begin{subfigure}{0.50\textwidth}
\begin{lstlisting}[firstline=19, lastline=42]
void relational_wrapper(int x1,int x2){
  int tmp1 = 0;
  int tmp2 = 0; 
  tmp1 = x1 + 1; // inlined f1(x1)
  tmp2 = x2 + 1; // inlined f1(x2)
/*@ assert x1 < x2 ==> tmp1 < tmp2; */
}

/*@ axiomatic Relational_axiom{
    logic int f1_acsl(int x);
    lemma Relational_lemma: \forall int x,y; 
      x < y ==> f1_acsl(x) < f1_acsl(y);
    }
*/

/*@ assigns \nothing;
  behavior Relational_behavior:
  ensures \result == f1_acsl(\old(x));
*/
int f1(int x){
  return x + 1;
}

... // similar f${}$or f2
\end{lstlisting}
    \vspace{-1mm}
  \end{subfigure}
  \vspace{-3mm}
  \lstset{basicstyle=\normalsize\ttfamily,}
  \caption{(a) Two  monotonic functions \lstinline{f1,f2} with three relational properties (file \lstinline{f.c}), 
and (b) excerpt of their transformation by \toolname for deductive verification}
  \vspace{-3mm}
  \label{fig:ex1}
\end{figure}

\smallskip\noindent
\textbf{Preprocessing and Proof Status Propagation.}
Since this new syntax is not supported by classic deductive verification tools, 
we have designed a code transformation, inspired by self-composition, 
allowing the user to prove the property 
with one of these tools. 

We illustrate the transformation for function \lstinline{f1} and property
\lstinline|R1| (see Figure \ref{fig:ex1}a).
%
%
The transformation result (Figure \ref{fig:ex1}b) consists of three parts. 
First, a new function, called \emph{wrapper}, is generated.
The wrapper function is inspired by the workaround proposed in \cite{BishopBC13}
and self-composition.
It inlines the function calls occurring in the relational property,
records their results in local variables and states an assertion equivalent
to the relational property (lines 1--7 in Figure \ref{fig:ex1}b).
The proof of such an assertion is possible with a classic deductive verification tool (\Wp can prove it in this example). 

However, a wrapper function is not sufficient if we need to use the relational 
property as a hypothesis in other proofs and to make their
support fully automatic and transparent for the user.
For this purpose, we generate an axiomatic definition (cf. \lstinline{axiomatic} section at lines 9--14)
to give a logical reformulation of the relational property as a lemma (cf. lines 11--12). 
This logical formulation can be used in subsequent proofs
(as we illustrate below).
Lemmas can refer to several function calls, but only for \emph{logic} functions.
Therefore, a logic counterpart (with \lstinline{_acsl} suffix) is declared 
for each C function involved in a relational property (cf. line 10). 
The ACSL function is partially specified {\it via} lemmas corresponding to
the relational properties of the original C function.
Note that the correspondence between \lstinline|f| and \lstinline|f_acsl| 
implies that
\lstinline|f| does not access global memory 
(neither for writing nor for reading). Indeed, since \lstinline|f_acsl| is a
pure logic function, it has no side effect and 
its result only depends on its parameters. 
Extending our approach for this case can rely on 
\lstinline'assigns...\from...' clauses, similarly to what is
proposed in~\cite{DBLP:journals/sttt/CuoqMPP11}, for adding to
\lstinline|f_acsl|
parameters representing the relevant parts of the program state.
This extension is left as future work.

Finally, to create a bridge between the C function and its logic counterpart,
we add a postcondition (an  \lstinline{ensures} clause, placed in a separate \lstinline{behavior} for readibility) 
to state that they always return the same result (cf. line 18 relating 
\lstinline{f1} and  \lstinline{f1_acsl}).

\begin{figure}[tb]
  \lstset{basicstyle=\scriptsize\ttfamily,mathescape=true}
  \begin{subfigure}{0.49\textwidth}
\begin{lstlisting}[firstline=1, lastline=24]
/*@ relational R1: \forall int x1,x2; 
  x1 < x2 ==> \call(f1,x1) < \call(f1,x2);
*/
int f1(int x);

/*@ relational R2: \forall int x1,x2; 
  x1 < x2 ==> \call(f2,x1) < \call(f2,x2);
*/
int f2(int x);

/*@ relational Rg: \forall int x1,x2; 
  x1 < x2 ==> \call(g,x1) < \call(g,x2);
*/
int g(int x){
  return f1(x)+f2(x);
}

/*@ relational Rh: \forall int x1,x2; 
  x1 < x2 ==> \call(h,x1) < \call(h,x2);
*/
int h(int x){
  return f1(f2(x));
}
\end{lstlisting}
    \vspace{-1mm}
  \end{subfigure}
  \hspace{2mm}
  \begin{subfigure}{0.49\textwidth}
\begin{lstlisting}[firstline=25, lastline=47]
void relational_wrapper(int x1,int x2){
  int tmp1 = 0;
  int tmp2 = 0;
  tmp1 = f1(x1) + f2(x1); // g(x1)
  tmp2 = f1(x2) + f2(x2); // g(x2)
/*@ assert x1 < x2 ==> tmp1 < tmp2;*/
}

/*@ axiomatic Relational_axiom{
    logic int g_acsl(int x);
    lemma relational_lemma: \forall int x,y; 
    x < y ==> g_acsl(x) < g_acsl(y);
    }
*/

/*@ behavior Relational_behavior:
  ensures \result == g_acsl(\old(x));
*/
int g(int x){
  return f1(x)+f2(x);
}

... // similar f${}$or h
\end{lstlisting}
    \vspace{-1mm}
  \end{subfigure}
  \vspace{-3mm}
  \lstset{basicstyle=\normalsize\ttfamily,}
  \caption{(a) Two  monotonic functions \lstinline{g,h} with two relational properties, 
and (b) extract of their transformation by \toolname for deductive verification}
  \vspace{-3mm}
  \label{fig:ex2}
\end{figure}

To make the proposed solution as transparent as possible for the user
and to ensure automatic propagation of proof statuses in the \framac property
database \cite{combining-analysis}, two additional 
rules are necessary.
First, the postconditions making the link between C functions and their 
associated logic counterparts are always supposed valid (so the clause of line 18
is declared as valid).
Second, the logic reformulation of a relational property in a 
lemma (lines 11--12) is declared valid\footnote{Technically, 
a special ``valid under condition'' status is used in this case in \framac.}
as soon as the assertion (line 6) at the end of the wrapper function is proved.

\vspace{-2mm}
\subsection{Implementation and Illustrative Examples} 
\vspace{-2mm}
\label{subsec:implementation}
\smallskip\noindent
\textbf{Implementation.}
A proof-of-concept implementation
 of the proposed technique has been realized in a \framac plugin 
\toolname (Relational Property Prover). 
\toolname works like a preprocessor for \Wp:
after its execution on a project containing relational properties, 
the proof on the generated code proceeds like any other proof with \Wp \cite{Frama-C}: 
proof obligations are generated and can be either discharged automatically
by automatic theorem provers (e.g.
Alt-Ergo, 
CVC4, 
Z3\footnote{See, resp., \url{https://alt-ergo.ocamlpro.com},
\url{http://cvc4.cs.nyu.edu},
\url{https://z3.codeplex.com/}})
or proven interactively (e.g. in Coq\footnote{See \url{http://coq.inria.fr/}}).

Thanks to the proposed code transformation no significant modification was required
in \framac and \Wp. \toolname currently supports 
relational properties of the form 
\vspace{-4mm}
\begin{center}
\lstset{basicstyle=\scriptsize\ttfamily,}
\begin{scriptsize}
$\forall$ \ \lstinline{<args1>}, \dots, $\forall$ \lstinline{<argsN>}, \\
$P($ \lstinline'<args1>', \dots,\lstinline'<argsN>', \lstinline'\call(f_1,<args1>)', \dots, \lstinline{\call(f_N,<argsN>)}$)$
\end{scriptsize}
\lstset{basicstyle=\normalsize\ttfamily,}
\end{center}
\vspace{-3mm}
%
for an arbitrary predicate $P$ invoking $N\ge 1$ calls of non-recursive functions without side effects
and complex data structures. 





\smallskip\noindent
\textbf{Illustrative Examples.}
After preprocessing with \toolname, \framac/\Wp automatically validates 
properties \lstinline{R1-R3} of Fig. \ref{fig:ex1}a  
by proving the assertions in the generated wrapper functions and by propagating proof statuses. 

To show how relational properties can be used in another proof, consider properties \lstinline{Rg,Rh}
of Figure \ref{fig:ex2}a for slightly more complex functions (inspired by \cite{BishopBC13})
whose proof needs to use properties \lstinline{R1,R2}. 
Thanks to their reformulation as lemmas and to the link between logic and C functions
(cf. lines 11--12, 18 of Figure \ref{fig:ex1}b for \lstinline{f1}), 
\Wp automatically proves the assertion at line 6 of Figure \ref{fig:ex2}b
and validates property \lstinline{Rg} as proven. 
The proof for \lstinline{Rh} is similar.

Notice that in examples of Figure \ref{fig:ex2},
functions  \lstinline{f1,f2} can be undefined 
since only their (relational) specification is required, which is suitable
for specification of library or hardware-provided functions
that cannot be specified without relational properties.

The \toolname tool has also been successfully tested on several other examples
such as cryptographic properties like \smash{$\mathrm{Decrypt}(\mathrm{Encrypt}(Msg,PrivKey),PubKey)=Msg$}, squeeze lemma condition (i.e. \smash{$\forall x, \  f_1(x) \leq f_2(x) \leq f_3(x)$}), 
median function properties (e.g. \smash{$ \forall a,b,c, \ \mathrm{Med}(a,b,c) = \mathrm{Med}(a,c,b)$}),
properties of  determinant for matrices of order 2 and 3 (e.g. \smash{$\mathrm{det}(A) = \mathrm{det}(A^\intercal)$}), matrix equations like 
$(A+B)^\intercal = A^\intercal+B^\intercal$, etc.
Some of them include loops 
whose loop invariants are automatically transferred by \toolname 
into the wrapper 
function to make possible its automatic proof.
\vspace{-4mm}
\section{Conclusion and Future Work}
\vspace{-3mm}
\label{sec:conclusion}
We proposed a novel technique for specification and proof of relational properties for C programs in \framac.
We implemented it in a \framac plugin \toolname and illustrated its capacity 
to treat a large range of examples coming from various industrial and academic projects
that were suffering from the impossibility to express relational properties.
One benefit of this approach is its capacity to rely on sound and mature
verification tools like \framac/\Wp, thus allowing for automatic or interactive proof from the specified code. 
Thanks to an elegant transformation into auxiliary C code and logic definitions accompanied by 
a property status propagation, the user can treat complex relational properties and observe the
results in a convenient and fully automatic manner.
Another key benefit is that this approach is  
suitable for verification of programs relying on library or hardware-provided functions 
whose source code is not available.

Future work includes extending the tool to support 
complex data structures and functions with side-effects,
support of recursive functions, studying other variants of
generated code (e.g. avoiding function inlining in some cases), 
as well as further experiments on real-life programs.




\begin{small}
\smallskip
\noindent
\textit{Acknowledgment.} 
Part of the research work leading to these results has received funding
for DEWI project (www.dewi-project.eu) from the ARTEMIS
Joint Under\-ta\-king under grant agreement No. 621353, 
and for the S3P project from French DGE and BPIFrance.
\end{small}

\vspace{-5mm}
\bibliographystyle{splncs03}
\bibliography{biblio}



\end{document}